\documentclass[twocolumn,aip,jcp,10pt,superscriptaddress,longbibliography]{revtex4-1}

\usepackage{amssymb}
\usepackage{amsmath}
\usepackage{multirow}
\usepackage{diagbox}
\usepackage{glossaries}
\usepackage{graphicx}
\usepackage{dcolumn}
\usepackage{bm}
\usepackage{subfigure}
\usepackage[usenames,dvipsnames,svgnames]{xcolor}
\usepackage{hyperref}
\hypersetup{
pdfnewwindow=true,      
colorlinks=true,        
linkcolor=Blue,         
citecolor=Blue,         
filecolor=Blue,         
urlcolor=Blue           
}
\usepackage{listings} 
\usepackage{siunitx}

\newacronym{ltc}{LTC}{lattice thermal conductivity}
\newacronym{dft}{DFT}{density functional theory}
\newacronym{md}{MD}{molecular dynamics}
\newacronym{emp}{EMP}{empirical potential}
\newacronym{mlp}{MLP}{machine-learned potential}
\newacronym{ml}{ML}{machine learning }
\newacronym{nep}{NEP}{neuroevolution potential}
\newacronym{nn}{NN}{neural network}
\newacronym{snes}{SNES}{separable natural evolution strategy}
\newacronym{mbd}{MBD}{many-body dispersion}
\newacronym{rmse}{RMSE}{root mean square error}
\newacronym{hnemd}{HNEMD}{homogeneous non-equilibrium molecular dynamics}
\newacronym{nemd}{NEMD}{non-equilibrium molecular dynamics}
\newacronym{emd}{EMD}{equilibrium molecular dynamics}
\newacronym{shc}{SHC}{spectral heat current}
\newacronym{mfp}{MFP}{mean free path}
\newacronym{bte}{BTE}{Boltzmann transport equation}
\newacronym{ald}{ALD}{anharmonic lattice dynamics}
\newacronym{aimd}{AIMD}{\emph{ab initio} molecular dynamics}
\newacronym{ann}{ANN}{artificial neural network}
\newacronym{dp}{DP}{deep potential}
\newacronym{gap}{GAP}{Gaussian approximation potential}
\newacronym{hcacf}{HCACF}{heat current autocorrelation function}
\newacronym{nist}{NIST}{national institute of standards and technology}
\newacronym{nqe}{NQE}{nuclear quantum effect}
\newacronym{rdf}{RDF}{radial distribution function}
\newacronym{scan}{SCAN}{strongly constrained and appropriately normed}
\newacronym{vdos}{VDOS}{vibrational density of states}
\newacronym{adf}{ADF}{angular distribution function}
\newacronym{ace}{ACE}{atomic cluster expansion}
\newacronym{3d}{3D}{three dimensional}
\newacronym{2d}{2D}{two dimensional}

\DeclareSIUnit\angstrom{\text{Å}}
\DeclareSIUnit{\atom}{atom}
\DeclareSIUnit{\step}{step}
\DeclareSIUnit{\atomstepsecond}{\atom\step\per\second}

\begin{document}

\title{Dissimilar thermal transport properties in $\kappa$-Ga$_2$O$_3$ and $\beta$-Ga$_2$O$_3$ revealed by machine-learning homogeneous nonequilibrium molecular dynamics simulations}

\author{Xiaonan Wang}
\affiliation{School of Science, Harbin Institute of Technology, Shenzhen, 518055, P. R. China}

\author{Jinfeng Yang}
\affiliation{School of Science, Harbin Institute of Technology, Shenzhen, 518055, P. R. China}

\author{Penghua Ying}
\email{hityingph@163.com}
\affiliation{School of Science, Harbin Institute of Technology, Shenzhen, 518055, P. R. China}

\author{Zheyong Fan}
 \affiliation{College of Physical Science and Technology, Bohai University, Jinzhou 121013, P. R. China}

\author{Jin Zhang}
 \affiliation{School of Science, Harbin Institute of Technology, Shenzhen, 518055, P. R. China}
 
 \author{Huarui Sun}
 \email{huarui.sun@hit.edu.cn}
\affiliation{School of Science, Harbin Institute of Technology, Shenzhen, 518055, P. R. China}
\affiliation{Ministry of Industry and Information Technology Key Laboratory of Micro-Nano Optoelectronic Information System, Harbin Institute of Technology, Shenzhen, 518055, P. R. China.}

\date{\today}

\begin{abstract}
The lattice thermal conductivity (LTC) of Ga$_2$O$_3$ is an important property due to the challenge in the thermal management of high-power devices. We develop machine-learned neuroevolution potentials for single-crystalline $\beta$-Ga$_2$O$_3$ and $\kappa$-Ga$_2$O$_3$, and apply them to perform homogeneous nonequilibrium molecular dynamics simulations to predict their LTCs. The LTC of $\beta$-Ga$_2$O$_3$ was determined to be \SI{10.3 \pm 0.2}{\watt\per\meter\per\kelvin}, \SI{19.9 \pm 0.2}{\watt\per\meter\per\kelvin}, and \SI{12.6 \pm 0.2}{\watt\per\meter\per\kelvin} along [100], [010], and [001], respectively, aligning with previous experimental measurements. For the first time, we predict the LTC of $\kappa$-Ga$_2$O$_3$ along [100], [010], and [001] to be \SI{4.5 \pm 0.0}{\watt\per\meter\per\kelvin}, \SI{3.9 \pm 0.0}{\watt\per\meter\per\kelvin}, and \SI{4.0 \pm 0.1}{\watt\per\meter\per\kelvin}, respectively, showing a nearly isotropic thermal transport property. The reduced LTC of $\kappa$-Ga$_2$O$_3$ versus $\beta$-Ga$_2$O$_3$ stems from its restricted low-frequency phonons up to \SI{5}{\tera\hertz}. Furthermore, we find that the $\beta$ phase exhibits a typical temperature dependence slightly stronger than $\sim T^{-1}$, whereas the $\kappa$ phase shows a weaker temperature dependence, ranging from $\sim T^{-0.5}$ to $\sim T^{-0.7}$. 
\end{abstract}

\maketitle

\section{Introduction\label{intro}}
 Ultra-wide bandgap semiconductors, such as Ga$_2$O$_3$, diamond, AlN, etc., have also become the focus of attention materials for next-generation electronics and optoelectronics. Owing to a bandgap of about \SI{5}{\electronvolt}, exceptional breakdown electrical field, and cost-effective production, Ga$_2$O$_3$ offers considerable promise for ultra-high power devices applications.\cite{pearton2018review, zhang2022ultra, ueda1997anisotropy} However, the \gls{ltc} of Ga$_2$O$_3$ is subpar, leading to pronounced heat dissipation issues in certain semiconductor devices.\cite{pearton2018review}  Understanding the phonon thermal transport in Ga$_2$O$_3$ is crucial for its practical applications. 


The Ga$_2$O$_3$ crystal actually exists in five distinct phases: $\alpha$, $\beta$, $\gamma$, $\sigma$, and $\varepsilon$ (sometimes referred to as $\kappa$). Of these, the $\beta$ phase (space group C2/m) is the most stable one, which has been extensively explored for applications in deep-ultraviolet transparent conductive electrodes,\cite{orita2000deep} solar blind detectors,\cite{2006APL,zhou2022flexible} high-performance field effect transistors,\cite{hwang2014high} Schottky rectifiers \cite{2018JAP} and high temperature gas sensors.\cite{pearton2018review} In recent years, further efforts have been made to overcome the poor thermal stability and immature synthesis methods of other phases. Hexagonal crystal $\varepsilon$-Ga$_2$O$_3$ (space group P63mc) was reported to be the second most stable phase obtained at low temperatures \cite{playford2013structures, zhuo2017beta}, whereas subsequent studies claimed that the crystal structure of the polycrystalline form at low temperatures has an orthorhombic structure at the nanoscale (\SIrange[range-phrase = --]{5}{10}{\nano\meter}), named $\kappa$-Ga$_2$O$_3$ (space group Pna21).\cite{cora2017real, janzen2021comprehensive} The strong polarization in $\kappa$-Ga$_2$O$_3$ is a prominent feature that $\beta$-Ga$_2$O$_3$ does not possess, which may benefit potential device applications. For example, $\kappa$-Ga$_2$O$_3$ can form heterojunctions with other semiconductors such as GaN or AlN, and its polarization was utilized to regulate interface transport while effectively alleviating thermal problems. \cite{chen2022band, krishna2022band} Considering the distinct crystal structures of $\beta$ and $\kappa$-Ga$_2$O$_3$, a comparative understanding of the \gls{ltc} for both phases is essential for their device thermal design, particularly as the $\kappa$ phase remains unexplored.

Currently, growing large-scale single-crystalline $\kappa$-Ga$_2$O$_3$ is a challenging endeavor experimentally, and assessing its \gls{ltc} is anticipated to be even more demanding. From a computational perspective, its \gls{ltc} can be predicted using atomistic simulation techniques, such as \gls{md} simulations and the combination of \gls{bte} with the \gls{ald} method. While atomic interactions can be derived from either empirical potentials or quantum mechanical \gls{dft} calculations, intricate crystals like $\kappa$-Ga$_2$O$_3$, which possess 40 atoms in their primitive cell, introduce substantial hurdles - accuracy concerns for empirical potentials and computational overhead for \gls{dft}. Recently, \glspl{mlp}-based large-scale \gls{md} simulations have been demonstrated to be a reliable approach to calculate the \gls{ltc} of complex crystals including amorphous silicon \cite{wang2023quantum}, amorphous HfO$_2$ \cite{zhang2023prb},  amorphous silica \cite{liang2023mechanisms}, metal-organic frameworks \cite{ying2023sub}, and violet phosphorene \cite{ying2023variable}, which can account for phonon anharmonicity to arbitrary order.
Several \glspl{mlp} for Ga$_2$O$_3$ have been developed, addressing both the perfect bulk system \cite{Liu2020,Li2020} and more intricate, disordered structures \cite{zhao2023complex,Liu2023}. 

In this work, we apply the \gls{nep} framework \cite{fan2021neuroevolution, fan2022improving, fan2022jcp} to develop two \glspl{mlp} on demand for Ga$_2$O$_3$ against quantum-mechanical \gls{dft} calculations, one for $\beta$ phase and one for $\kappa$ phase. We choose the \gls{nep} approach here, because this method has been demonstrated to be highly efficient  \cite{fan2022jcp}. We apply the developed \glspl{nep} to perform extensive \gls{hnemd} simulations to investigate the \gls{ltc} of the two phases of Ga$_2$O$_3$. Our results show that the \gls{ltc} of the $\kappa$ phase is much lower than that of the $\beta$ phase, and exhibits low anisotropy and weak temperature dependence.

\section{Methods\label{meth}}
\subsection{Structural model and DFT calculations}
\label{section:dft}
All \gls{dft} calculations were performed using projected augmented wave method\cite{blochl1994projector} with Perdew-Burke-Ernzerhof functional of generalized gradient approximation \cite{perdew1996generalized} implemented in the \textsc{VASP} package \cite{Kresse1996PRB, Kresse1999PRB}. The kinetic energy cutoff was set to \SI{520}{\electronvolt}, and the convergence value for the total energy was \SI{1e-6}{\electronvolt}. The stopping criteria for structural optimizations were that the maximum residual Hellmann-Feynman force on atoms was less than \SI{1e-3}{\electronvolt\per\angstrom}.  A \numproduct{10x10x6} and \numproduct{10x6x6} $\Gamma$-centered k-point grid was employed for the primitive cell of $\beta$-Ga$_2$O$_3$ and $\kappa$-Ga$_2$O$_3$ respectively. Additionally, to validate the accuracy of the \gls{mlp}, the phonon dispersions of $\beta$-Ga$_2$O$_3$ and $\kappa$-Ga$_2$O$_3$ were further calculated on \numproduct{2x2x2} and \numproduct{2x1x1} supercells, respectively, using the density functional perturbation theory together with the \textsc{Phonopy} package \cite{togo2015first}. 

\begin{figure}[htb]
\begin{center}
\includegraphics[width=1\columnwidth]{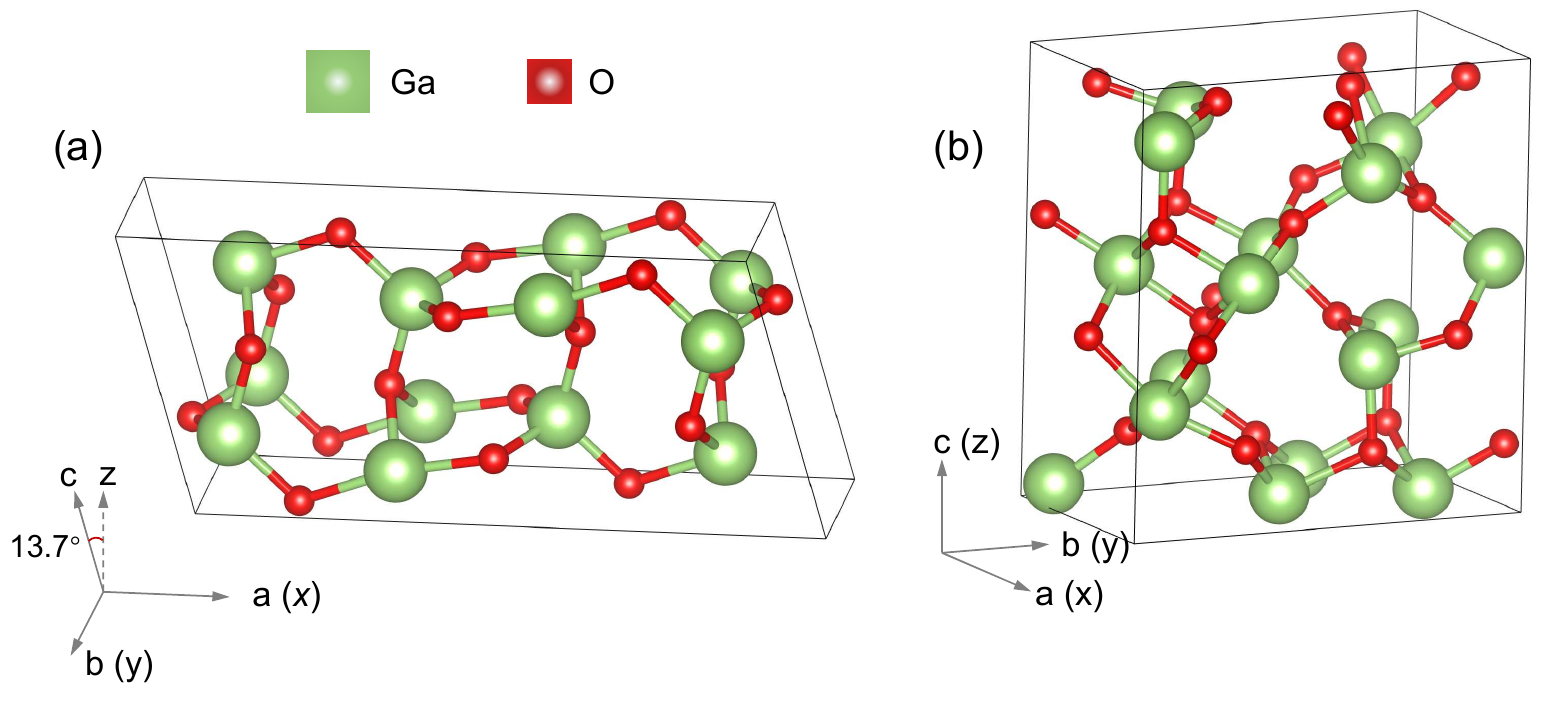}
\caption{\label{fig:structure} Crystal structure of the conventional unit cell of (a) $\beta$-Ga$_2$O$_3$ and (b) $\kappa$-Ga$_2$O$_3$.}
\end{center}
\end{figure}

Meanwhile, according to widely used conventions, we have optimized the corresponding conventional unit cell as the initial cell for the \gls{md} simulation. The $\kappa$-Ga$_2$O$_3$ lattice structure remains unchanged, but the number of atoms in the conventional unit cell of $\beta$-Ga$_2$O$_3$ is higher than that in the primitive cell. The calculated lattice constants of $\beta$-Ga$_2$O$_3$ are $a$ = \SI{12.468}{\angstrom}, $b$ = \SI{3.087}{\angstrom} and $c$ = \SI{5.716}{\angstrom}, while these values for $\kappa$-Ga$_2$O$_3$ are $a$ = \SI{5.074}{\angstrom}, $b$ = \SI{8.703}{\angstrom} and $c$ = \SI{9.309}{\angstrom}, which are close to reported values \cite{janzen2021comprehensive, aahman1996reinvestigation}. As shown in \autoref{fig:structure}, the conventional unit cell of $\beta$-Ga$_2$O$_3$ has 20 atoms, while $\kappa$-Ga$_2$O$_3$ has a relatively large number of 40 atoms. In the calculations of the reference datasets for \gls{nep} training, the \numproduct{2x2x2} and \numproduct{2x1x1} supercells were used for $\beta$-Ga$_2$O$_3$ and $\kappa$-Ga$_2$O$_3$, respectively, with \numproduct{1x4x2} and \numproduct{3x3x3} k-point grids.

\subsection{The NEP model training}
 We used the identical method to construct the reference data sets for the $\beta$ and $\kappa$ phases of Ga$_2$O$_3$. The reference structures were obtained by \gls{aimd} simulations and random perturbations. For each phase, the \gls{aimd} simulations were run under an isothermal ensemble with temperature increasing linearly from \SIrange{10}{1000}{\kelvin} for 10,000 steps and a time step of \SI{1}{\femto\second}. We uniformly extracted 1,000 structures from \gls{aimd} simulations for each case. For random perturbations, 200 structures were generated with random cell deformations ranging from -4\% to 4\% and atomic displacements less than \SI{0.1}{\angstrom}, all based on optimized structures. For both phases, our total data set has 1,200 structures including energy, atomic forces, and virials obtained from \gls{dft} calculations as outlined in \autoref{section:dft}. We randomly divided the total data set into a training set of 1,000 structures and a test set of 200 structures.

After obtaining the training set and test set, we applied the third generation of the \gls{nep} framework \cite{ fan2022jcp} implemented in \textsc{GPUMD} (version 3.5) to train the \gls{mlp} for $\beta$ and $\kappa$ phases of Ga$_2$O$_3$, which were denoted as \gls{nep}-$\beta$ and \gls{nep}-$\kappa$, respectively. \gls{nep} used a feedforward \gls{nn} to represent the site energy with  \gls{ace}\cite{drautz2019atomic}-like descriptor components including radial and angular terms. The parameters of \gls{nep} models were optimized using the \gls{snes}  \cite{Schaul2011}, with the loss function defined as a weighted sum over the \gls{rmse} values of energy, atomic force and virial.

We used the same hyperparameters for these two \gls{nep} models. After extensive testing, the selected hyperparameters were determined as follows: the radial and angular cutoffs are $r^R_{\rm c}$ = \SI{8}{\angstrom} and $r^A_{\rm c}$ = \SI{4}{\angstrom}, respectively. The number of radial and angular descriptor components are $n^R_{\rm max}+1=9$ and $n^A_{\rm max}+1=9$, respectively. For angular parameters, we have $l^{3b}_{\rm max}=4$ for three-body and $l^{4b}_{\rm max}=2$ for four-body terms, respectively. The number of neurons in the hidden layer of the \gls{nn} is 50. The size of the population is $N_{\rm pop}=50$ and the number of generations is $N_{\rm gen}=5\times 10^5$. The weights of energy, force, and virial \glspl{rmse} in the loss function were set to 1.0, 1.0, and 0.1, respectively. 

\subsection{Thermal conductivity calculations}

We performed \gls{md} simulation using the \textsc{GPUMD} package (version 3.5) \cite{fan2017cpc, fan2022jcp} to calculate the \glspl{ltc} for the two phases of  Ga$_2$O$_3$. Based on the \gls{hnemd} method, the \gls{ltc} can be calculated from the relation \cite{evans1982homogeneous, Fan2019PRB} 
\begin{equation}
\label{equation:kappa}
\frac{\langle J^{\mu}(t)\rangle_{\rm ne}}{TV} = \sum_{\nu} \kappa^{\mu\nu}  F^{\nu}_{\rm e},
\end{equation}
where $\kappa^{\mu\nu}$ is the thermal conductivity tensor, $T$ is the system temperature and $V$ is the system volume. The non-equilibrium heat current $\langle \bm{J} \rangle_{\rm ne}$ is induced by the external driving force $\bm{F}_{i}^{\rm ext}$ related to a driving-force parameter $\bm{F}_{\rm e}$ with the dimension of inverse length \cite{gabourie2021spectral}
\begin{equation}
\label{equation:Fe}
\boldsymbol{F}_{i}^{\rm ext}
=  \bm{F}_{\rm e} \cdot \mathbf{W}_i.
\end{equation}
Here, $\mathbf{W}_i$ is the virial tensor of atom $i$. The magnitude $F_{\rm e}$ of the driving-force parameter we used for both phases at different temperatures are small enough to keep the system within the linear-response regime.

All \gls{md} simulations were conducted at the target temperature using the Nose-Hoover chain method \cite{tuckerman2023statistical} with a time step of \SI{1.0}{\femto\second}. Initially, the simulation was run for \SI{0.1}{\nano\second} in the isothermal-isobaric ensemble, and subsequently in the isothermal ensemble for \SI{1.0}{\nano\second} to achieve equilibrium. Afterwards, \gls{hnemd} simulations were performed in the isothermal ensemble for \SI{10.0}{\nano\second} to calculate the running \glspl{ltc}. We employed \numproduct{4x16x8} and \numproduct{9x5x5} supercells for $\beta$-Ga$_2$O$_3$ and $\kappa$-Ga$_2$O$_3$, respectively, containing 10240 and 9000 atoms. The simulation sizes were validated to be sufficiently large to eliminate finite-size effects. For each case, the predicted \gls{ltc} was calculated as the average of five independent simulations, and the corresponding standard error was also estimated.

\begin{figure*}[htb]
\begin{center}
\includegraphics[width=2\columnwidth]{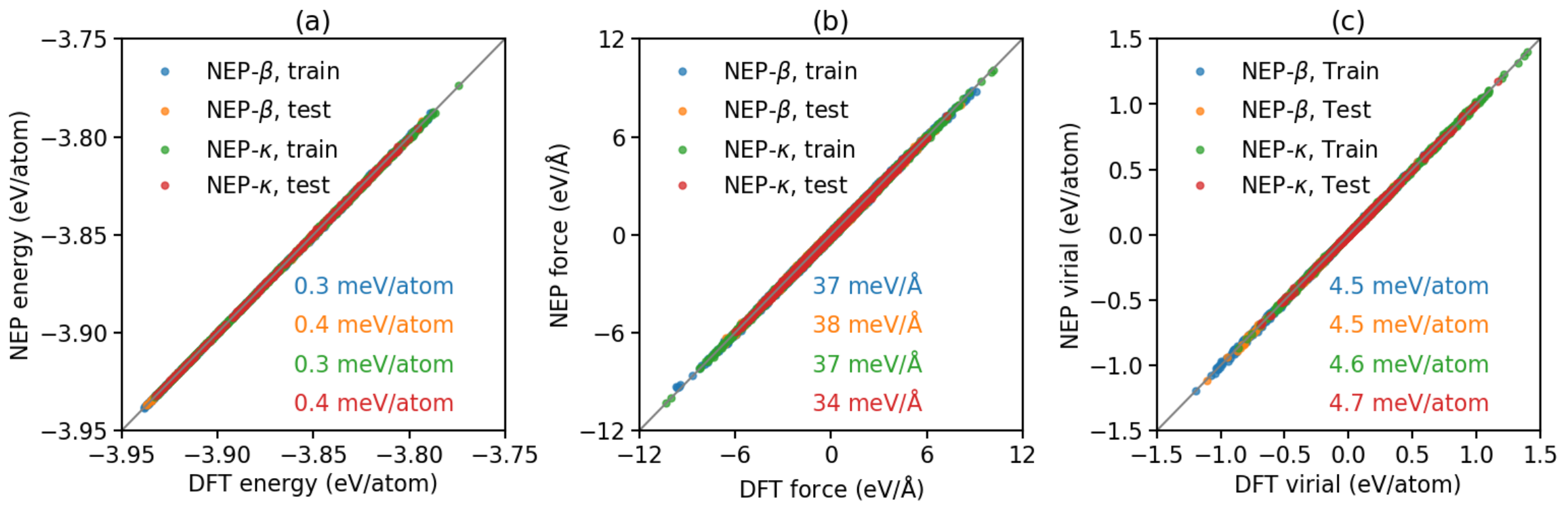}
\caption{\label{fig:rmse} The \gls{nep} predictions of energy, force, and virial for train and test sets against the \gls{dft} reference values.}
\end{center}
\end{figure*}

\section{Results and discussion}

\subsection{Validation of the NEP models}

\autoref{fig:rmse}(a)-(c) show a comparison between the predicted energy, force, and virial values obtained from the \gls{nep} models and the corresponding 
\gls{dft} reference values for both the training and test sets of the two phases. In all scenarios, the \glspl{rmse} for energy, force, and 
virial are below \SI{0.5}{\meV\per\atom}, \SI{40}{\meV\per\angstrom}, and \SI{5}{\meV\per\atom}, respectively. These results demonstrate a very high accuracy of our \gls{nep} models. In addition to  their high accuracy, our \glspl{nep} are remarkably efficient. For a system comprising 10,000 atoms, our \glspl{nep} can attain a computational speed of approximately \SI{1.5e+06}{} atom-step per second in the \textsc{GPUMD} package using a single GeForce RTX 3090 GPU card. This efficiency enabled us to perform extensive simulations to characterize the \glspl{ltc} of the Ga$_2$O$_3$ crystals.
 
\begin{figure}[htb]
\begin{center}
\includegraphics[width=\columnwidth]{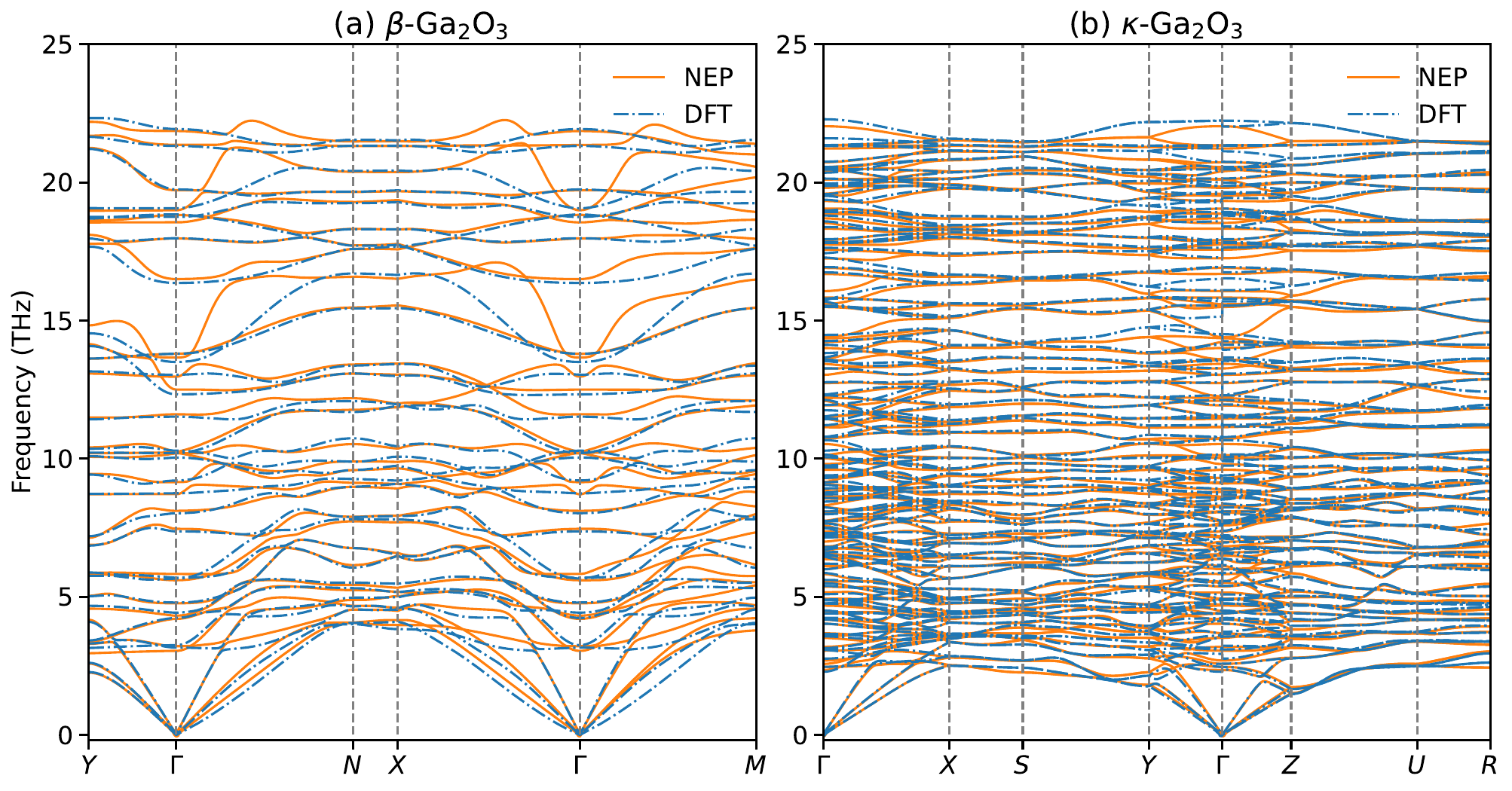}
\caption{\label{fig:phonon} The phonon dispersions of (a) $\beta$-Ga$_2$O$_3$ and (b) $\kappa$-Ga$_2$O$_3$ predicted by \gls{nep} and \gls{dft}.}
\end{center}
\end{figure}

\begin{figure}[htb]
\begin{center}
\includegraphics[width=\columnwidth]{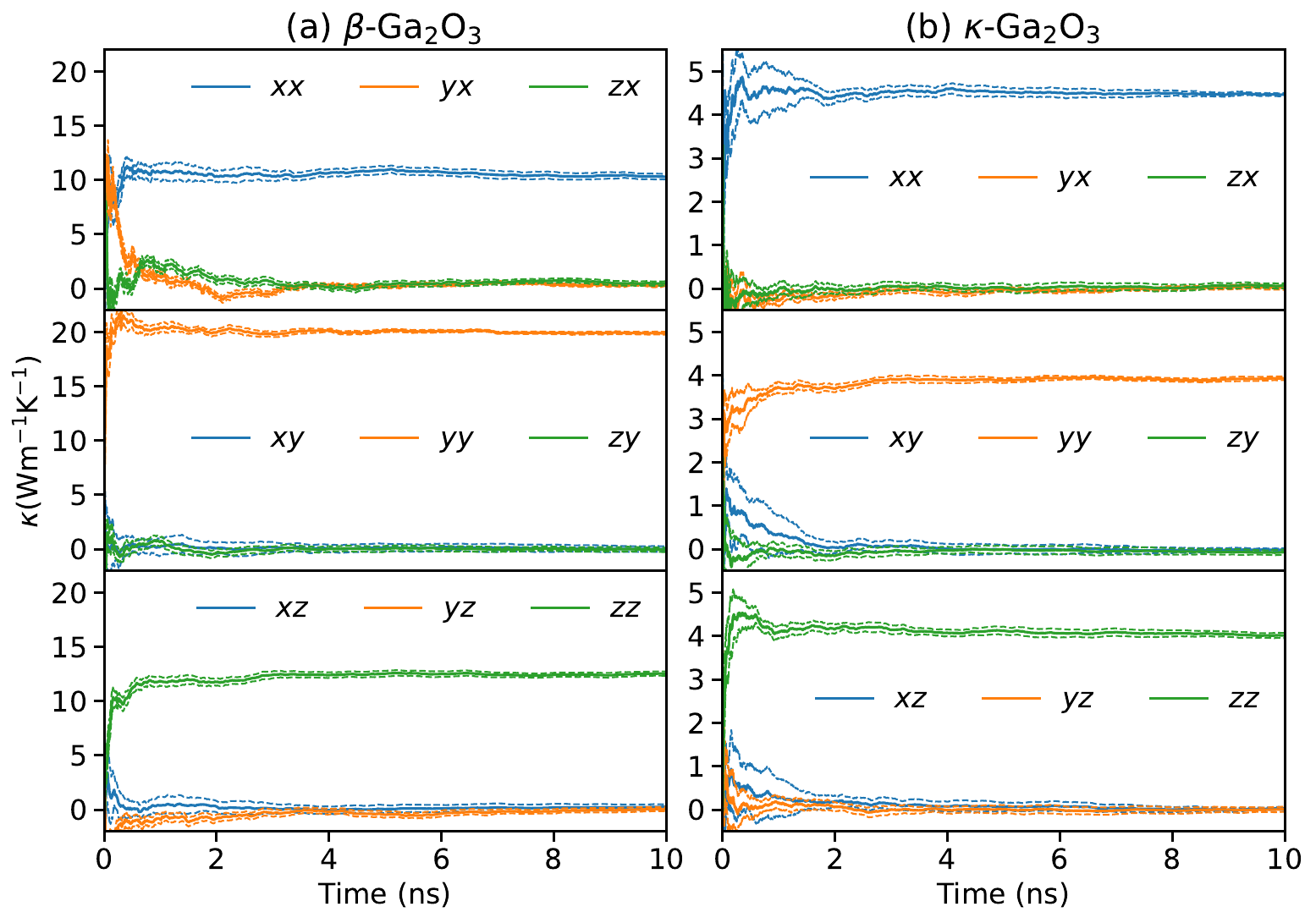}
\caption{\label{fig:kappa} Running \gls{ltc} tensors from \gls{hnemd} simulations at \SI{300}{\kelvin} for: (a) $\beta$- and (b) $\kappa$-Ga$_2$O$_3$. Driving forces for top, middle, and bottom are along the $x$, $y$, and $z$ directions, respectively.}
\end{center}
\end{figure}


\begin{table}[htb]
\setlength{\abovecaptionskip}{1cm}  
\caption{\label{table:tensor}The predicted \gls{ltc} tensor (\SI{}{\watt\per\meter\per\kelvin}) of (a) $\beta$-Ga$_2$O$_3$ and (b) $\kappa$-Ga$_2$O$_3$ from \gls{hnemd} simulations at \SI{300}{\kelvin}.}
\centering
\begin{tabular}{lll}
\hline
\gls{ltc} components & $\beta$-Ga$_2$O$_3$  & $\kappa$-Ga$_2$O$_3$ \\
\hline
$\kappa_{xx}$ & 10.3 (0.2) & 4.5 (0.0) \\
$\kappa_{yy}$ & 19.9 (0.2) & 3.9 (0.0) \\
$\kappa_{zz}$ & 12.5 (0.2) & 4.0 (0.1) \\
$\kappa_{xy}$ & 0.2 (0.2) & 0.0 (0.1) \\
$\kappa_{xz}$ & 0.4 (0.2) & 0.0 (0.1) \\
$\kappa_{yz}$ & 0.0 (0.1) & 0.0 (0.1) \\
\hline
\label{table:ltc tensor}
\end{tabular}
\end{table}

\begin{figure*}[htb]
\begin{center}
\includegraphics[width=2\columnwidth]{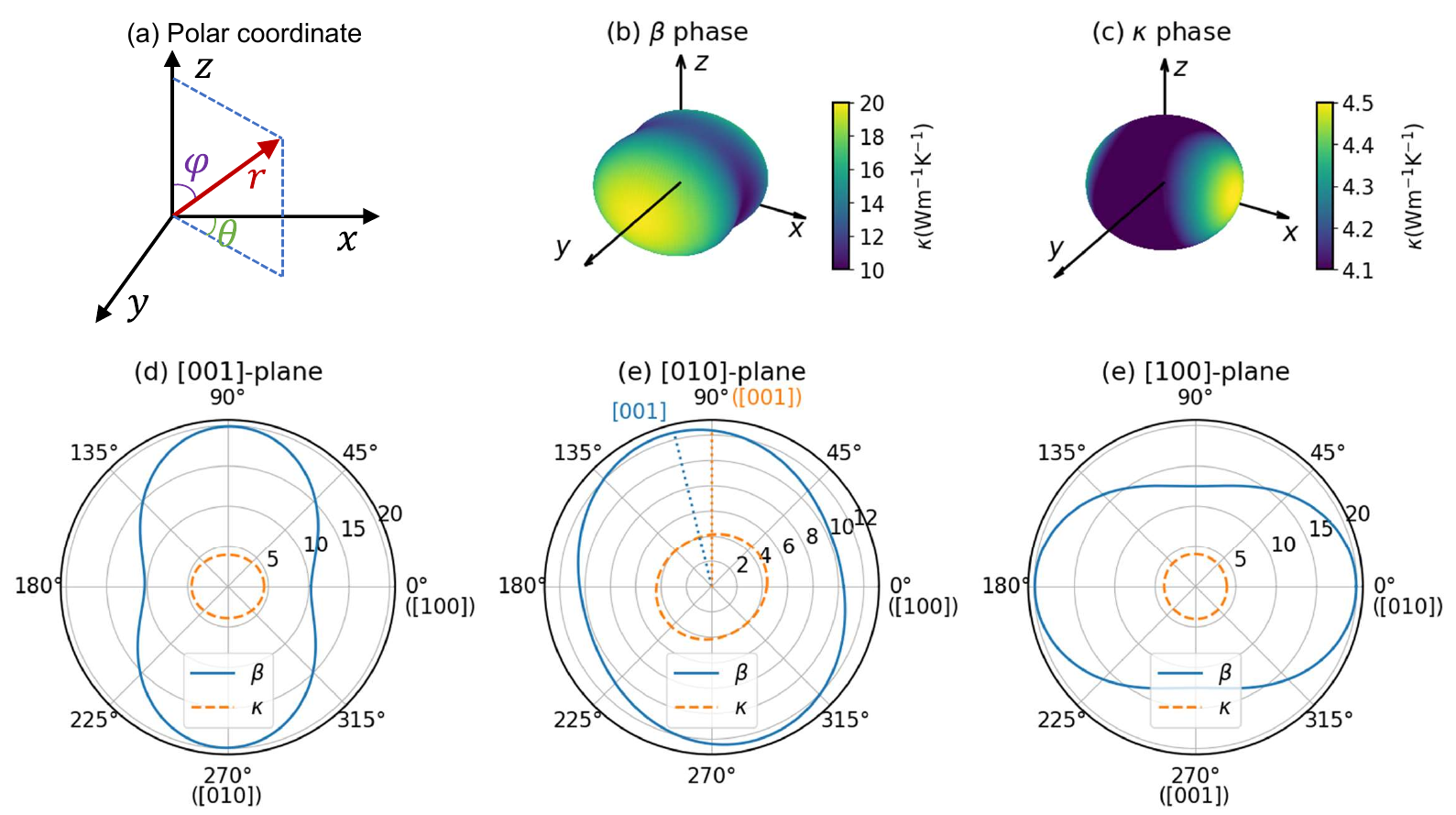}
\caption{\label{fig:sphere} (a) The polar coordination used for the transformation of the \gls{ltc} tensor. (b) and (c) The \gls{3d} distributions of the \gls{ltc} for $\beta$-Ga$_2$O$_3$ and $\kappa$-Ga$_2$O$_3$. (d), (e), and (f) The corresponding \gls{2d} projections onto the [100], [010], and [001]-planes, respectively. All \glspl{ltc} are in units of \SI{}{\watt\per\meter\per\kelvin}.}
\end{center}
\end{figure*}

\subsection{Phonon dispersions}

To assess the reliability of our \gls{nep} models in capturing the phonon transport properties of $\beta$-Ga$_2$O$_3$ and $\kappa$-Ga$_2$O$_3$, we compare the calculated phonon dispersions using both \gls{nep} and \gls{dft} methods, as shown in \autoref{fig:phonon}. It can be seen that for both phases, the acoustic branches predicted by \gls{nep} and \gls{dft} are very close, while the optical branches of $\beta$ phase show deviations, especially at high frequencies. Because the theoretical calculation finds that the high-frequency optical branches ($\textgreater \SI{10}{\THz}$) contribute minimally to \gls{ltc}, \cite{Yan2018pccp} which is confirmed by subsequent calculations of the spectrally decomposed \gls{ltc} in this work. Thus, we believe that the \gls{nep} models can reliably predict the \gls{ltc} of the two phases of Ga$_2$O$_3$.

\subsection{Thermal conductivity}
After confirming the accuracy of the \gls{nep} models, we apply them to calculate \glspl{ltc} using the \gls{hnemd} method.
\autoref{fig:kappa} presents the \glspl{ltc} for both 
$\beta$- and $\kappa$-Ga$_2$O$_3$ using the \gls{hnemd} method at \SI{300}{\kelvin}. Notably, \autoref{fig:structure}(a) reveals that $\beta$-Ga$_2$O$_3$ possesses a monoclinic crystal structure, 
suggesting the potential for non-zero values in the 
off-diagonal elements in the \gls{ltc} tensor. Fortunately, the \gls{hnemd} approach (see \autoref{equation:kappa}) allows us to fully determine the \gls{ltc} tensor as listed in \autoref{table:tensor}, by applying the external driving force along three orthogonal axes sequentially. As anticipated, the $\beta$ phase exhibits minor couplings between the $x$ and $z$ directions. In contrast, the off-diagonal elements of the \gls{ltc} tensor for the $\kappa$ phase are zero, attributed to its orthorhombic crystal structure.

Based on the polar coordinates as shown in \autoref{fig:sphere}(a), we can determine the \gls{ltc} along any crystal direction:\cite{Jiang2018} 
\begin{equation}
\label{equation:polar}
\kappa(\theta, \phi) = 
\boldsymbol{\alpha} \begin{pmatrix}
\kappa_{xx} & \kappa_{xy} & \kappa_{xz} \\
\kappa_{xy} & \kappa_{yy} & \kappa_{yz} \\
\kappa_{xz} & \kappa_{yz} & \kappa_{zz}
\end{pmatrix} \boldsymbol{\alpha}^{T},
\end{equation}
with
\begin{equation}
\boldsymbol{\alpha} = 
\begin{pmatrix}
\sin{\phi}\cos{\theta}, & \sin{\phi}\sin{\theta}, & \cos{\phi}
\end{pmatrix}.
\end{equation}
\autoref{fig:sphere}(b) and (c) show the \gls{3d} distribution of \gls{ltc} for the $\beta$ and $\kappa$ phases, respectively. The corresponding \gls{2d} projection in the [001], [010], and [100]-planes are presented in \autoref{fig:sphere}(d)-(f), respectively. 

For the $\beta$ phase, the \gls{ltc} is highest along the [010] direction with \SI{19.9 \pm 0.2}{\watt\per\meter\per\kelvin}, followed by [001] with \SI{12.6 \pm 0.2}{\watt\per\meter\per\kelvin} (see the blue dotted line in \autoref{fig:sphere}(e)), and [100] at \SI{10.3 \pm 0.2}{\watt\per\meter\per\kelvin}. As shown in \autoref{table:comparisons}, our \gls{hnemd} results are consistent with previous experimental results \cite{Jiang2018, Guo2015apl} using time-domain thermoreflectance and theoretical predictions based on the \gls{bte}-\gls{ald} method \cite{santia2015lattice, Liu2020,Yan2018pccp, ChenJPCC2023} or the \gls{emd} method \cite{Li2020}. This further demonstrates the reliability of our \gls{hnemd} approach based on machine-learned \gls{nep} in characterizing the \gls{ltc} of Ga$_2$O$_3$ crystals. 

\begin{table}[htb]
\setlength{\abovecaptionskip}{1cm}  
\caption{The \gls{ltc} (\SI{}{\watt\per\meter\per\kelvin}) of $\beta$-Ga$_2$O$_3$ at \SI{300}{\kelvin} predicted by our \gls{hnemd} simulations and previously reported values. The values in parentheses denote the standard errors.}
\centering
\begin{tabular}{llll}
\hline
Method & [100] & [010] & [001]\\
\hline
Current \gls{hnemd} & 10.3 (0.2) & 19.9 (0.2) & 12.6 (0.2)\\
Experiment \cite{Jiang2018} & 9.5 (1.8)  & 22.5 (2.5) & 13.3 (1.8)\\
Experiment \cite{Guo2015apl}  & 10.9 (1.0) & 27.0 (2.0) & 15.0\\
\gls{emd}\cite{Li2020} & 10.7 & 20.8 & 12.6\\
\gls{bte}-\gls{ald}\cite{Liu2020} & 13.9 & 24.8 & 19.8\\
\gls{bte}-\gls{ald}\cite{santia2015lattice}  &16.1 & 21.5 & 21.2\\
\gls{bte}-\gls{ald}\cite{Yan2018pccp} & 12.7 & 20.0 & 17.8\\
\gls{bte}-\gls{ald}\cite{ChenJPCC2023} & 10.02 & 23.74 & 12\\

\hline
\label{table:comparisons}
\end{tabular}
\end{table}

To our knowledge, the \gls{ltc} of $\kappa$-Ga$_2$O$_3$ has not been investigated before. Here, we predict the \gls{ltc} for $\kappa$-Ga$_2$O$_3$ to be \SI{4.5 \pm 0.0}{\watt\per\meter\per\kelvin}, \SI{3.9 \pm 0.0}{\watt\per\meter\per\kelvin}, and \SI{4.0 \pm 0.1}{\watt\per\meter\per\kelvin} along the [100], [010], and [001] directions, respectively. These values are about one-fifth to half of those of $\beta$-Ga$_2$O$_3$. While the $\beta$ phase displays a significant \gls{ltc} anisotropy with an anisotropy index of 1.94 (the maximum-to-minimum ratio of the \gls{ltc}, see spatial distribution in \autoref{fig:sphere}(b)), the $\kappa$ phase exhibits a nearly isotropic \gls{ltc} pattern with an anisotropy index of 1.15. (see \autoref{fig:sphere}(c)).

To elucidate the disparities in the \gls{ltc} observed between the two phases, we decomposed the \gls{ltc} as a function of phonon frequency: \cite{Fan2019PRB}
\begin{equation}
\kappa^{\mu\nu}(\omega)=\frac{2}{V T F_{\rm{e}}^{\nu}} \int_{-\infty}^{\infty} \mathrm{d} t \mathrm{e}^{i \omega t} K^{\rm \mu}(t).
\label{equation:shc}
\end{equation}
Here $\bm{K}(t)=\sum_i\left\langle\mathbf{W}_i(0) \cdot \bm{v}_i(t)\right\rangle$ is the virial-velocity correlation function \cite{gabourie2021spectral}, in which $\mathbf{W}_i$ and $\bm{v}_i$ are the virial tensor and the velocity of atom $i$, respectively.

\begin{figure}[htb]
\begin{center}
\includegraphics[width=\columnwidth]{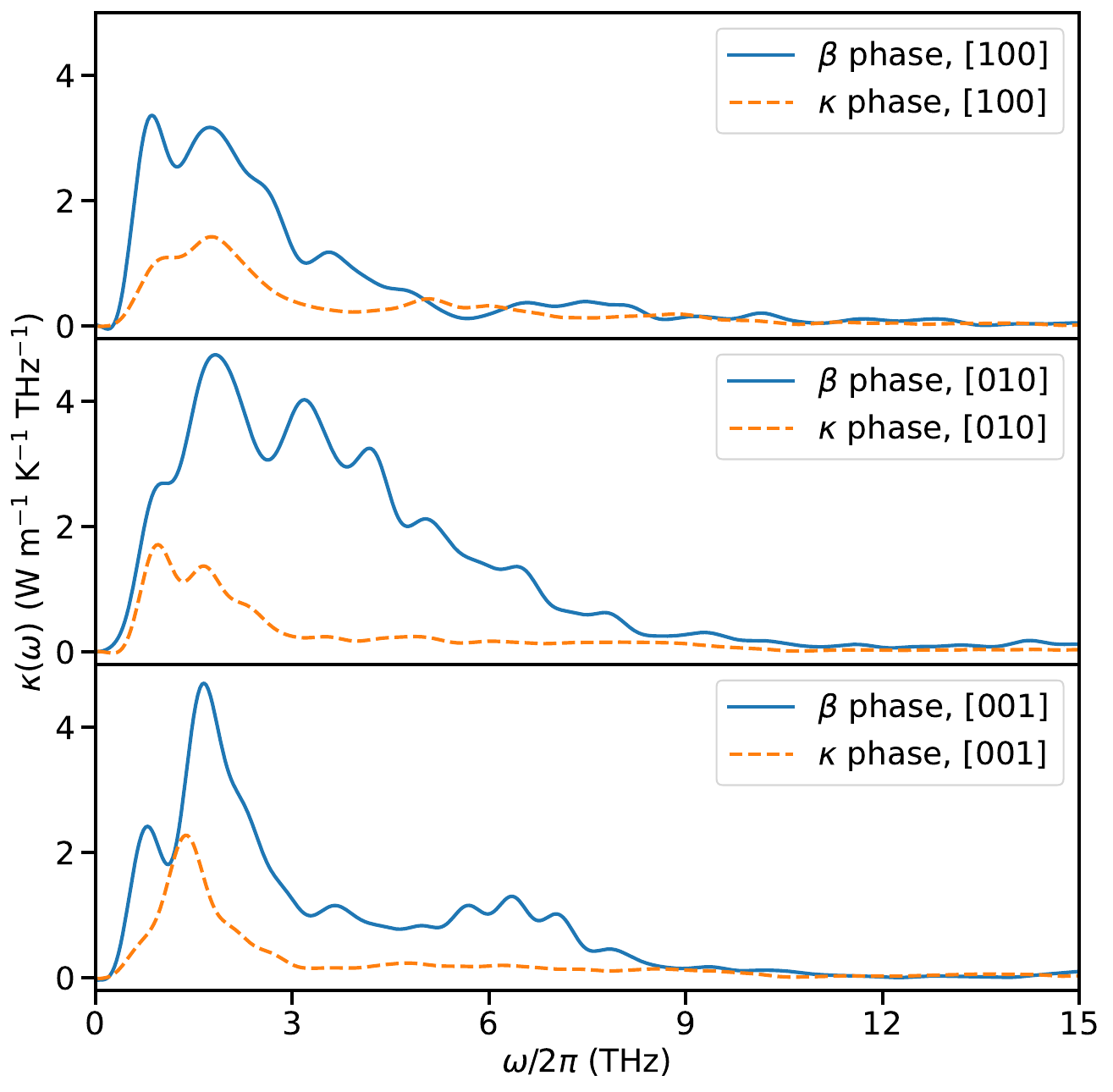}
\caption{\label{fig:shc} The spectrally decomposed \gls{ltc} of (a) $\beta$-Ga$_2$O$_3$ and (b) $\kappa$-Ga$_2$O$_3$ as a function of phonon frequency.}
\end{center}
\end{figure}

\autoref{fig:shc} presents the spectrally decomposed \gls{ltc} of $\beta$-Ga$_2$O$_3$ and $\kappa$-Ga$_2$O$_3$ calculated from \autoref{equation:shc}. Taken as a whole, only phonon modes with frequencies in the \SI{0}-\SI{10} {\tera\hertz} range are really involved in thermal transport. It also can be found that, although in both phases the low-frequency acoustic phonons contribute significantly to the \gls{ltc}, the $\beta$ phase has a much broader distribution of phonon frequencies contributing to its \gls{ltc} than the $\kappa$ phase. This is also related to the the phonon dispersions as shown in \autoref{fig:phonon}. The branches of the $\kappa$ phase with a phonon frequency in the range of \SIrange{5}{10}{\tera\hertz} are much flatter than those of the $\beta$ phase, leading to much lower phonon group velocities. Moreover, in the $\kappa$ phase, we observe significant overlaps between multiple bands for phonon frequencies greater than \SI{3}{\tera\hertz}, indicating the presence of multiple scattering channels, a phenomenon previously observed in Violet phosphorene \cite{ying2023variable}.

\begin{figure}[htb]
\begin{center}
\includegraphics[width=\columnwidth]{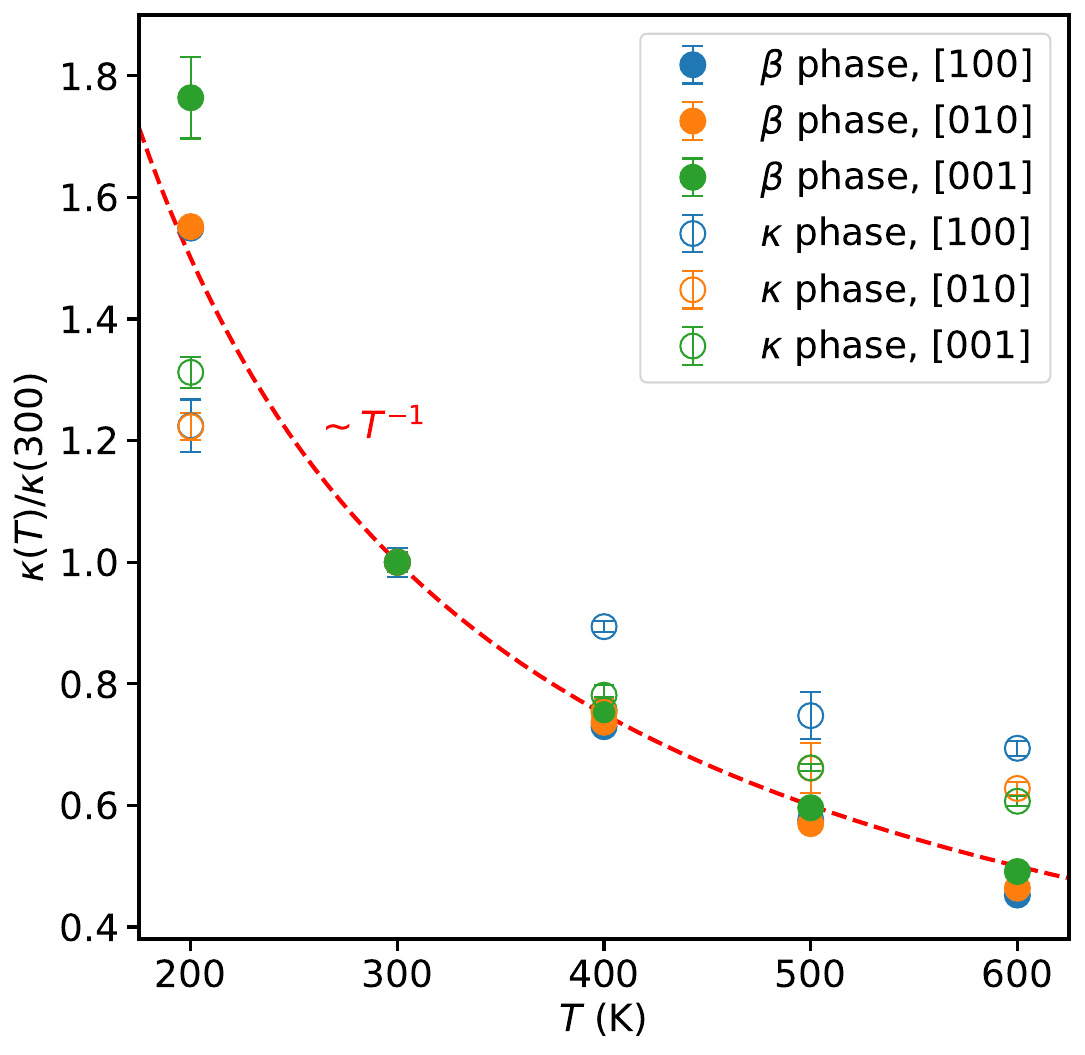}
\caption{\label{fig:temp} \gls{ltc} of $\beta$ and $\kappa$ phases of Ga$_2$O$_3$ as a function of temperature $T$ along three directions. All the \glspl{ltc} are normalized by their values at \SI{300}{\kelvin}, respectively.}
\end{center}
\end{figure}

Furthermore,  \autoref{fig:temp} shows the temperature-dependent \gls{ltc} for $\beta$- and $\kappa$-Ga$_2$O$_3$. In the temperature range of \SIrange{200}{600}{\kelvin}, the \glspl{ltc} of both phases decrease with increasing temperature. The \gls{ltc} of the $\beta$ phase exhibits a temperature dependence slightly stronger than $\sim T^{-1}$, as typical for systems dominated by three-phonon scattering processes \cite{2013LindsayPhysRevB}. However, the \gls{ltc} of $\kappa$-Ga$_2$O$_3$ shows a clearly weaker temperature dependence, which is $\sim T^{-0.5}$ along the [100] direction and $\sim T^{-0.7}$ along the the [010] and [001] directions, suggesting a significant impact of higher-order anharmonic phonon scatterings attributable to its complex crystal structures \cite{2019LindsayJAP}.

\section{Conclusions}
In summary, we have developed machine-learned \gls{nep} models trained against quantum-mechanical \gls{dft} data for the $\beta$ and $\kappa$ phases of Ga$_2$O$_3$, which have been demonstrated to be accurate and efficient in predicting energy, atomic forces, virial, and phonon dispersions in both phases. Based on large-scale \gls{hnemd} simulations, we reached a consistent prediction with previous experimental measurements for the $\beta$ phase, and predicted the \gls{ltc} of the $\kappa$ phase for the first time. We found that the $\kappa$ phase has a much lower \gls{ltc} than the $\beta$ phase, due to its phonon frequency contributions being limited to below \SI{5}{\tera\hertz} compared to the \SIrange{0}{10}{\tera\hertz} range of the $\beta$ phase. Furthermore, the $\kappa$ phase exhibits an almost isotropic spatial distribution with an anisotropy index of 1.15 at \SI{300}{\kelvin}, in contrast to the pronounced anisotropy index of 1.94 for the $\beta$ phase. We also examined the temperature dependence of the \gls{ltc} in the two phases, and found that the \gls{ltc} of the $\beta$ phase follows a temperature dependence slightly stronger than $\sim T^{-1}$, whereas the $\kappa$ phase shows a weaker temperature dependence from $\sim T^{-0.5}$ to $\sim T^{-0.7}$, indicating a significant effect of high-order anharmonicity distinct as in low-\gls{ltc} materials. Our wok demonstrates that the machine-learned \gls{nep}-driven \gls{hnemd} simulations can reliably and effectively characterize phonon thermal transport properties for complex crystals such as $\kappa$-Ga$_2$O$_3$, and thus we expect that this approach can be used to explore the \glspl{ltc} of other phases of Ga$_2$O$_3$ as well.

\begin{acknowledgments}
This work was supported by the Key-Area Research and Development Program of Guangdong Province (Grant No.2020B010169002), the Guangdong Special Support Program (Grant No.2021TQ06C953), and the Science and Technology Planning Project of Shenzhen Municipality (Grant No. JCYJ20190806142614541). 
\end{acknowledgments}

\vspace{0.5cm} 

\noindent{\textbf{Conflict of Interest}}

The authors have no conflicts to disclose.

\vspace{0.5cm} 

\noindent{\textbf{Data availability}}
Complete input and output files for the \gls{nep} training of $\beta$-Ga$_2$O$_3$ and $\kappa$-Ga$_2$O$_3$ crystals are freely available at \url{https://gitlab.com/brucefan1983/nep-data}.The source code and documentation for  \textsc{GPUMD} are available
at \url{https://github.com/brucefan1983/GPUMD} and \url{https://gpumd.org}, respectively.

\end{document}